\documentclass[11pt]{article}
\usepackage{amsmath}
\usepackage{amsfonts}
\usepackage[dvips]{epsfig}

 \advance \textwidth by -2in
 \advance \textheight by -3in
\def\##1{{\bf #1}}
\def\=#1{\underline{\underline{#1}}}

\def\+#1{\underline{\bf #1}}
\def\*#1{\underline{\underline{\bf #1}}}

\def\eps{\epsilon}
\def\epso{\epsilon_{\scriptscriptstyle 0}}
\def\muo{\mu_{\scriptscriptstyle 0}}
\def\ko{k_{\scriptscriptstyle 0}}

\def\.{\mbox{ \tiny{$^\bullet$} }}

\def\lec{\left\{}
\def\ric{\right\}}

\def\c#1{\cite{#1}}

\begin{document}

\noindent {\bf SIMULTANEOUS NEGATIVE-- AND POSITIVE--PHASE--VELOCITY\\
PROPAGATION IN AN ISOTROPIC CHIRAL MEDIUM  }
\vskip 0.2cm

\noindent  {\bf Tom G. Mackay$^1$ and Akhlesh Lakhtakia$^2$} \vskip
0.2cm

\noindent {\sf $^1$ School of Mathematics\\
\noindent University of Edinburgh\\
\noindent Edinburgh EH9 3JZ, United Kingdom} \vskip 0.4cm

\noindent {\sf $^2$ CATMAS~---~Computational \& Theoretical Materials Sciences Group \\
\noindent Department of Engineering Science \& Mechanics\\
\noindent 212 Earth \& Engineering Sciences Building\\
\noindent Pennsylvania State University, University Park, PA
16802--6812} \vskip 0.4cm

\

\noindent {\bf ABSTRACT:} The phase velocity of plane waves
propagating in an isotropic chiral medium can be simultaneously
positive for left--circularly polarized light and negative for
right--circularly polarized light (or vice versa). The constitutive
parameter regimes supporting this phenomenon are established for
nondissipative and dissipative mediums. The boundary between
positive and negative phase velocity is  characterized by
infinite phase velocity.

\vskip 0.2cm \noindent {\bf Keywords:} {\em negative refraction,
positive refraction, infinite phase velocity} \vskip 0.4cm

\vspace{10mm}

\noindent{\bf 1. INTRODUCTION}

An isotropic chiral medium is characterized in the frequency domain by the Tellegen constitutive relations
\begin{eqnarray}
&&\#D = \epso\eps_r\,\#E +i\sqrt{\epso\muo}\,\kappa \,\#H\,,\\
&&\#B = \muo\mu_r\,\#H -i\sqrt{\epso\muo}\,\kappa \,\#E\,,
\end{eqnarray}
wherein $\eps_r$, $\mu_r$, and $\kappa$ are frequency--dependent complex--valued scalars,
whereas $\epso$ and $\muo$ are the permittivity and permeability of
free space (i.e., vacuum).

Electromagnetic waves in this medium are best described using left-- and
right--handed  Beltrami
fields \c{Beltrami}
\begin{eqnarray}
&&{\#Q}_L = \#E + i \frac{\sqrt{\muo}}{\sqrt{\epso}}\,\frac{\sqrt{\mu_r}}{\sqrt{\eps_r}}\,\#H\,,\\
&&{\#Q}_R = \#E - i \frac{\sqrt{\muo}}{\sqrt{\epso}}\,\frac{\sqrt{\mu_r}}{\sqrt{\eps_r}}\,\#H\,,
\end{eqnarray}
such that
\begin{eqnarray}
&&\nabla\times{\#Q}_L = k_L\,{\#Q}_L\,,\quad k_L= \ko(\sqrt{\eps_r}\sqrt{\mu_r}+\kappa)\,,\\
&&\nabla\times{\#Q}_R = -\,k_R\,{\#Q}_R\,,\quad k_R= \ko(\sqrt{\eps_r}\sqrt{\mu_r}-\kappa)\,.\end{eqnarray}

Examining plane--wave propagation in an isotropic chiral medium,
Mackay \c{Mackay} obtained the  conditions to delineate
negative--phase--velocity (NPV) and positive--phase--velocity (PPV)
propagation therein. Those conditions led to the following question:
Can an isotropic chiral medium simultaneously support NPV and PPV
propagation? If the answer is in the affirmative, an isotropic
chiral slab could behave like the so--called perfect lens for
right--circularly polarized (RCP) light from a certain source but
not for left--circularly polarized (LCP) light from that same source
(or vice versa) \c{Pendry}.

\noindent{\bf 2. ANALYSIS}

Let us begin with the nondissipative scenario wherein
 $\eps_r$, $\mu_r$, and $\kappa$ are all real--valued.
 The  restriction $\eps_r \mu_r > 0$ is imposed  to exclude  evanescent plane waves from consideration.
We have that
\begin{equation*}
\begin{array}{lcl}
\left| \kappa \right| < \left| \sqrt{\eps_r} \sqrt{\mu_r} \right| &
\Rightarrow & \left. \begin{array}{l} \mbox{LCP and RCP plane waves
are of the PPV type for $\eps_r > 0$}
\\
\mbox{LCP and RCP plane waves are of the NPV type for $\eps_r < 0$}
\end{array} \right\},
\end{array}
\end{equation*}
while
\begin{equation*}
\begin{array}{lcl}
 - \kappa > \left| \sqrt{\eps_r} \sqrt{\mu_r}
\right| & \Rightarrow & \left.
\begin{array}{l}
 \mbox{LCP plane waves are of the  NPV type}
\\ \mbox{RCP plane waves are of the PPV
type}
\end{array} \right\}
\end{array}
\end{equation*}
and
\begin{equation*}
\begin{array}{lcl}
 \kappa > \left| \sqrt{\eps_r} \sqrt{\mu_r} \right| & \Rightarrow &
\left.
\begin{array}{l}
 \mbox{LCP plane waves are of the  PPV type}
\\ \mbox{RCP plane waves are of the NPV
type}
\end{array} \right\}.
\end{array}
\end{equation*}
Thus, when
\begin{equation}
| \kappa | > \left| \sqrt{\eps_r} \sqrt{\mu_r} \right|\,,
\end{equation}
the nondissipative isotropic chiral
medium simultaneously supports both PPV and NPV
plane--wave propagation. The transition from PPV to NPV (or vice
versa) arises as the relevant wavenumber ($k_L$ or $k_R$) passes through zero. Therefore, the
PPV/NPV boundary is characterized by the magnitude of the phase velocity
becoming infinite, as has been reported previously for isotropic
dielectric--magnetic mediums \c{LM04}.

Next we turn to the dissipative scenario wherein
 $\eps_r$, $\mu_r$, and $\kappa$ are all complex--valued.
The wavenumbers lie in the upper half of the complex plane, as
dictated by
  the
Kramers--Kronig relations \c{BH}. Let us consider the regimes $\rho
> 0$ and $\rho < 0$, where\footnote{The operators
$\mbox{Re} \lec \. \ric$ and $\mbox{Im} \lec \. \ric$ deliver the
real and imaginary parts of their complex--valued arguments.}
\begin{equation}
\rho = \frac{\mbox{Re} \lec \eps_r \ric}{\mbox{Im} \lec \eps_r \ric}
+ \frac{\mbox{Re} \lec \mu_r \ric}{\mbox{Im} \lec \mu_r \ric}
\end{equation}
is the NPV parameter for isotropic dielectric--magnetic mediums
\c{DL04}. We find
\begin{equation*}
\begin{array}{lcl}
\left| \mbox{Re} \lec \kappa \ric \right| < \mbox{Re} \lec
\sqrt{\eps_r} \sqrt{\mu_r} \ric & \Rightarrow & \left.
\begin{array}{l}
\mbox{LCP and RCP plane waves are of the PPV type for $\rho > 0$}
\\
\mbox{LCP and RCP plane waves are of the NPV type for $\rho < 0$}
\end{array} \right\},
\end{array}
\end{equation*}
whereas
\begin{equation*}
\begin{array}{lcl}
 - \mbox{Re} \lec \kappa  \ric >  \mbox{Re} \lec \sqrt{\eps_r} \sqrt{\mu_r}
\ric & \Rightarrow & \left.
\begin{array}{l}
 \mbox{LCP plane waves are of the  NPV type}
\\ \mbox{RCP plane waves are of the PPV
type}
\end{array} \right\}
\end{array}
\end{equation*}
and
\begin{equation*}
\begin{array}{lcl}
  \mbox{Re} \lec \kappa \ric >  \mbox{Re} \lec \sqrt{\eps_r} \sqrt{\mu_r} \ric & \Rightarrow &
\left.
\begin{array}{l}
 \mbox{LCP plane waves are of the  PPV type}
\\ \mbox{RCP plane waves are of the NPV
type}
\end{array} \right\}.
\end{array}
\end{equation*}
 Therefore, the dissipative medium
simultaneously supports both PPV and NPV plane--wave propagation
when
\begin{equation}
\left| \mbox{Re} \lec \kappa \ric \right| > \mbox{Re} \lec
\sqrt{\eps_r} \sqrt{\mu_r} \ric\,.
\end{equation}
The PPV/NPV boundary occurs where
the real part of the relevant wavenumber passes through zero and the
magnitude of the phase
velocity correspondingly becomes unbounded \c{LM04}.

\noindent{\bf 3. CONCLUDING REMARKS}

When the magnitude of the real part of its relative
magnetoelectric parameter $\kappa$ is sufficiently
large relative to the magnitudes of the real parts of its relative permittivity
$\epsilon_r$ and relative
permeability $\mu_r$, an isotropic chiral medium can simultaneously support
both PPV and NPV plane--wave propagation, with different circular
polarization states. Large values of $\kappa$ are difficult to realize \c{Boh},
and perhaps the best option to simultaneously propagate NPV and PPV
plane waves in an isotropic chiral material would be to focus on realizing
$|\eps_r| < 1$ and $|\mu_r| < 1$ \c{LM06}.
 \vspace{10mm}

\noindent{\bf Acknowledgment:}  TGM is supported by a \emph{Royal
Society of Edinburgh/Scottish Executive Support Research
Fellowship}.

\end{document}